\documentclass[epsfig,floats,pre,twocolumn,preprintnumbers,floatfix]{revtex4-1}
\usepackage{graphicx}
\usepackage[latin2]{inputenc}
\usepackage{amsmath}
\usepackage{amssymb}
\usepackage{bm}
\usepackage{color}

\begin{document}
\title{Transient anomalous diffusion in run-and-tumble dynamics}
\author{M. Reza Shaebani}
\email{shaebani@lusi.uni-sb.de}
\author{Heiko Rieger}
\affiliation{Department of Theoretical Physics $\&$ Center for Biophysics, Saarland University, 66041 
Saarbr\"ucken, Germany}

\begin{abstract}
We study the stochastic dynamics of a particle with two distinct motility states. 
Each one is characterized by two parameters: one represents the average speed 
and the other represents the persistence quantifying the tendency to maintain 
the current direction of motion. We consider a run-and-tumble process, which is 
a combination of an active fast motility mode (persistent motion) and a passive 
slow mode (diffusion). Assuming stochastic transitions between the two motility 
states, we derive an analytical expression for the time evolution of the mean 
square displacement. The interplay of the key parameters and the initial 
conditions as for instance the probability of initially starting in the run 
or tumble state leads to a variety of transient regimes of anomalous transport 
on different time scales before approaching the asymptotic diffusive dynamics. 
We estimate the crossover time to the long-term diffusive regime 
and prove that the asymptotic diffusion constant is independent of initially 
starting in the run or tumble state. 
\end{abstract}

\maketitle
\section{Introduction}

Many transport processes in nature involve distinct motility states. Of 
particular interest is the run-and-tumble process, which consists of alternating 
phases of fast active and slow passive motion. Prominent examples are bacterial 
species that swim when their flagella form a bundle and synchronize their 
rotation. The bundle is disrupted and swimming stops when some of the flagella 
stochastically change their rotational direction. In the absence of rotating 
bundle, the bacterium moves diffusively until it manages to re-form the bundle 
and actively move forward again \citep{Berg04,Berg72}. The run-and-tumble 
dynamics is beneficial for bacteria as it allows them to react to the 
environmental changes by adjusting their average run time or speed 
\citep{Patteson15}, change their direction of motion, perform an efficient 
search \citep{Bartumeus08,Benichou05,Najafi18,Benichou11}, or optimize their 
navigation \citep{Wadhams04,Taktikos14}. 

Another example is the motion of molecular motors along cytoskeletal filaments. 
When motor proteins bind to filaments, they perform a number of steps until 
they randomly unbind and experience diffusion in the crowded cytoplasm. While 
the efficiency of long-distance cargo delivery requires high motor processivity 
(i.e.\ the tendency to continue the motion along the filament), the slow 
diffusive mode during unbinding periods is also vital for cellular functions 
which depend on the localization of the reactants \citep{Weigel11,Guigas08,
Golding06,Sereshki12}. The processivity of the motors (and thus the unbinding 
probability) depends on the type of motor and filament \citep{Ali07,
Shiroguchi07} and the presence of particular proteins or binding domains in 
the surrounding medium \citep{Vershinin07,Okada03,Culver-Hanlon06}. On the 
other hand other factors, such as cell crowding, may affect the binding 
probability. Therefore, the switching probabilities between active run and 
tumble states are generally asymmetric. By ignoring the microscopic details 
of stepping on filaments, coarse-grained random walk models have been employed 
to study the two-state dynamics of molecular motors \citep{Klumpp05,Lipowsky05,
Hafner16,Pinkoviezky13}. Dentritic immune cells also move persistently 
(migration phase) interrupted by slow phases for antigen uptake \citep{Chabaud15}. 
There have been many other locomotive patterns in biological and non-living 
systems investigated via models with distinct states of motility \citep{Bressloff13,
Hofling13,Angelani13,Soto14,Shaebani16,Theves13,Elgeti15,Thiel12}. For instance, 
the problem of searcher proteins finding a specific target site over a DNA strand 
has been studied by multi-state stochastic processes \citep{Berg81,Meroz09,Bauer12}.

The particle trajectories obtained from experiments often comprise a set 
of recorded positions of the particle, from which the successive directions 
of motion can be deduced. These directions are correlated on short time scales 
for active motions. However, the trajectory eventually gets randomized and 
the asymptotic dynamics is diffusive, with a diffusion constant $D_\text{asymp}$ 
that depends on the particle velocity and persistency \citep{Nossal74,
Sadjadi15}. One expects a similar long-term behavior for a mixture of run 
and tumble dynamics as well. The question arises how the transient short time 
dynamics, the crossover time to asymptotic diffusion, and $D_\text{asymp}$ 
depend on the run and tumble velocities and the switching probabilities 
between the two states. It is also unclear how the overall dynamics is 
influenced by the choice of the initial conditions of motion, like the 
probabilities to start either in the run or tumble state, which are parameters 
that can be extracted from experimental data.

Here, we present a two-state model for the run-and-tumble dynamics with 
spontaneous switchings between the states of motility. By deriving an 
analytical expression for the time evolution of the mean square displacement, 
we show how the interplay between the run and tumble velocities, the transition 
probabilities, and the initial conditions of motion leads to various 
anomalous transport regimes on short and intermediate time scales. We 
particularly clarify how the probability of starting from run or tumble 
state diversifies the transient anomalous regimes of motion, and verify 
that the long-term diffusion constant $D_\text{asymp}$ does not depend on 
the choice of the initial conditions of motion. 

\begin{figure*}[t]
\begin{center}
\includegraphics[width=12cm]{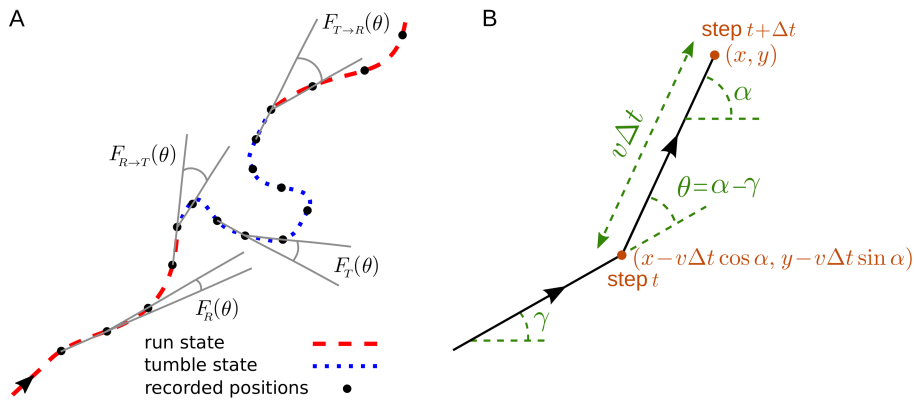}
\end{center}
\caption{(A) A sample trajectory with run-and-tumble dynamics. Typical 
turning angles for different types of turning-angle distributions introduced 
in the model are shown. (B) Trajectory of the walker during two successive 
steps.}
\label{Fig1}
\end{figure*}

\section{Model}

We develop a stochastic model for the run-and-tumble dynamics with spontaneous 
transitions between the motility states. We consider a two-state random walk 
in discrete time and continuous space with the following characteristics: The 
{\it run} phase is a persistent random walk with persistency $p$ and mean speed 
$v_{_\text{R}}$. The dynamics in the {\it tumble} phase is an ordinary diffusion 
with the mean speed $v_{_\text{T}}$. The asymmetric transition probabilities 
from run to tumble phase and vice versa are denoted, respectively, by 
$f_{_{_{\text{R}{\rightarrow}\text{T}}}}$ and $f_{_{_{\text{T}{\rightarrow}
\text{R}}}}$. As a result of constant transition probabilities, the run and 
tumble times are exponentially distributed in our model. This restriction can be 
relaxed by introducing time-dependent transition probabilities \citep{Shaebani19}. 
To characterize the persistency of the run phase, we use the probability 
distribution $F_{_\text{R}}(\theta)$ of directional changes along the trajectory 
in the run phase. The directional persistence can be characterized by the 
persistency parameter $p\,{=}\int_{-\pi}^{\pi}\text{d}\theta\;e^{\text{i}\theta}
F_{_\text{R}}(\theta)$, which leads to $p{=}\langle\cos\theta\rangle$ for symmetric 
distributions with respect to the arrival direction. Thus, $p$ ranges from $0$ for 
pure diffusion to $1$ for ballistic motion and reflects the average curvature of 
the run trajectories. Similarly, we define $F_{_\text{T}}(\theta)$ for the 
probability distribution of directional changes along the trajectory in the 
tumble phase, and $F_{_{_{\text{R}{\rightarrow}\text{T}}}}(\theta)$ and 
$F_{_{_{\text{T}{\rightarrow}\text{R}}}}(\theta)$ for the directional changes when 
switching between the states occurs [see Fig.~\ref{Fig1}(A)]. In the tumble phase 
(i.e.\ an ordinary diffusion), the probability distribution of directional changes 
is isotropically distributed ($F_{_\text{T}}(\theta){=}\frac{1}{2\pi}$), leading 
to a zero persistency.  

The run-and-tumble stochastic process can be described in discrete time by 
introducing the probability densities $P_{t}^{R}(x,y|\alpha)$ and $P_{t}^{T}
(x,y|\alpha)$ to find the particle at position $(x,y)$ arriving along the 
direction $\alpha$ at time $t$ in the run and tumble states, respectively. 
$\alpha$ is defined with respect to a given reference direction, as shown in 
Fig.~\ref{Fig1}(B). Denoting the time interval between consecutive recorded 
positions of the particle by $\Delta t$, the following set of master equations 
describe the dynamical evolution of the probability densities
\begin{equation}
\begin{array}{ll}
P_{\!_{t{+}\Delta t}}^{R}(x,y|\alpha) = \vspace{1mm} \\
\,(1{-}f_{_{_{\text{R}{\rightarrow}\text{T}}}}) 
\!\! \displaystyle\int_{\!{-}\pi}^{\pi} \!\!\!\! 
d\gamma F_{_\text{R}}(\alpha{-}\gamma) 
P_{t}^{R}(x{-}v_{_\text{R}}\Delta t\cos\alpha,y{-}
v_{_\text{R}}\Delta t\sin\alpha|\gamma) \vspace{1mm} \\
\,+ f_{_{_{\text{T}{\rightarrow}\text{R}}}} 
\!\! \displaystyle\int_{\!{-}\pi}^{\pi} \!\!\!\! 
d\gamma F_{_{_{\text{T}{\rightarrow}\text{R}}}}(\alpha{-}\gamma) 
P_{t}^{T}(x{-}v_{_\text{R}}\Delta t\cos\alpha,y{-}
v_{_\text{R}}\Delta t\sin\alpha|\gamma), \vspace{3mm} \\
\\
P_{\!_{t{+}\Delta t}}^{T}(x,y|\alpha) = \vspace{1mm} \\
\,(1{-}f_{_{_{\text{T}{\rightarrow}\text{R}}}}) 
\!\! \displaystyle\int_{\!{-}\pi}^{\pi} \!\!\!\! 
d\gamma F_{_\text{T}}(\alpha{-}\gamma) 
P_{t}^{T}(x{-}v_{_\text{T}}\Delta t\cos\alpha,y{-}
v_{_\text{T}}\Delta t\sin\alpha|\gamma) \vspace{1mm} \\
\,+ f_{_{_{\text{R}{\rightarrow}\text{T}}}} 
\!\! \displaystyle\int_{\!{-}\pi}^{\pi} \!\!\!\! 
d\gamma F_{_{_{\text{R}{\rightarrow}\text{T}}}} 
P_{t}^{R}(x{-}v_{_\text{T}}\Delta t\cos\alpha,y{-}
v_{_\text{T}}\Delta t\sin\alpha|\gamma).
\end{array}
\label{Eq:MasterEqs}
\end{equation}
Each of the two terms on the right-hand side of the equations 
represents the possibility of being in one of the two states 
in the previous time step (see Fig.~\ref{Fig1}(B) for the particle 
trajectory during two successive steps). The probability of 
starting the motion in the run or tumble phase is denoted by 
$P_0^R$ and $P_0^T$, respectively (with $P_0^T{=}1{-}P_0^R$). The 
change in the direction of motion $\theta{=}\alpha{-}\gamma$ with 
respect to the arrival direction is randomly chosen according to the 
turning-angle distribution $F_{_\text{R}}(\theta)$ or $F_{_\text{T}}
(\theta)$ in the run or tumble state, respectively. Both distributions 
are symmetric with respect to the arrival direction (i.e.\ left-right 
symmetric in 2D). We assume for simplicity that the directional change 
during the transition between the two states follows the turning-angle 
distribution of the new state, corresponding to $F_{_{_{\text{R}{
\rightarrow}\text{T}}}}(\theta){=}F_{_\text{T}}(\theta)$ and 
$F_{_{_{\text{T}{\rightarrow}\text{R}}}}(\theta){=}F_{_\text{R}}
(\theta)$. However, in general one should consider independent 
turning-angle distributions with non-zero mean for $F_{_{_{\text{R}{
\rightarrow}\text{T}}}}(\theta)$ and $F_{_{_{\text{T}{\rightarrow}
\text{R}}}}(\theta)$ as, for instance, a sharp change in the direction 
of motion of {\it E.\ coli} or {\it Bacillus Subtilis} when switching 
from tumbling to running is observed \citep{Berg04,Berg72,Najafi18}.

The total probability density $P_{t{+}\Delta t}(x,y|\alpha)$ to find the 
particle at position $(x,y)$ arriving along the direction $\alpha$ at 
time $t{+}\Delta t$ is given by $P_{t{+}\Delta t}(x,y|\alpha)=P_{t{+}
\Delta t}^{R}(x,y|\alpha)+P_{t{+}\Delta t}^{T}(x,y|\alpha)$. Using the 
Fourier transform of the probability density in each state $h$ ($h{\in}
\{R,T\}$), defined as
\begin{align}
P_{\!_{t{+}\Delta t}}^{h}(\boldsymbol{k}|m) \equiv \int_{-\pi}^{\pi} 
\!\!\!\! \text{d}\alpha\,e^{im\alpha} \int \!\! \text{d}y \int \!\! \text{d}x 
\,\, e^{i\boldsymbol{k}\cdot\boldsymbol r} P_{\!_{t{+}\Delta t}}^{h}(x,y|\alpha),
\label{Eq:Fourier-Run}
\end{align}
the Fourier transform of the total probability density is given by 
$P_{\!_{t{+}\Delta t}}(\boldsymbol{k}|m)=P_{\!_{t{+}\Delta t}}^{R}(
\boldsymbol{k}|m)+P_{\!_{t{+}\Delta t}}^{T}(\boldsymbol{k}|m)$, from which 
the moments of displacement can be calculated as
\begin{eqnarray}
\begin{aligned}
&\displaystyle\langle x^{j_{1}} y^{j_{2}} \rangle(t{+}\Delta t) 
\equiv \!\! \int \!\! \text{d}\alpha \!\! \int \!\! \text{d}y \!\! 
\int \!\! \text{d}x \,\, x^{j_{1}} y^{j_{2}}  P_{\!_{t{+}\Delta t}}
(x,y|\alpha)\\
&{=}(-i)^{j_{1}+j_{2}} \frac{\partial^{j_{1}+j_{2}} 
P_{\!_{t{+}\Delta t}}(k_{x},k_{y}|m=0)}{\partial k_{x}^{j_{1}}
\partial k_{y}^{j_{2}}} \Bigg|_{(k_{x},k_{y})=(0,0)}.
\end{aligned}
\label{Eq:Moments}
\end{eqnarray}
By means of a Fourier-$z$-transform technique, it is possible 
to solve the master equations~(\ref{Eq:MasterEqs}) to obtain 
the time evolution of the moments of displacement \citep{Sadjadi15,
Sadjadi08,Shaebani14}. Here we briefly explain the procedure 
to calculate the mean square displacement (MSD) as the main 
quantity of interest. From Eq.~(\ref{Eq:Moments}), the MSD is 
given as
\begin{equation}
\displaystyle\langle x^{2} \rangle(t{+}\Delta t) = 
(-i)^{2} \frac{\partial^{2} P_{\!_{t{+}\Delta t}}(k,\phi
{=}0|m{=}0)}{\partial k^{2}} \Bigg|_{k{=}0},
\label{Eq:MSD-1}
\end{equation}
where $(k,\phi)$ is the polar representation of $\boldsymbol{k}$. 
Assuming $F_{_{_{\text{R}{\rightarrow}\text{T}}}}(\theta){=}
F_{_\text{T}}(\theta){=}\frac{1}{2\pi}$ and $F_{_{_{\text{T}
{\rightarrow}\text{R}}}}(\theta){=}F_{_\text{R}}(\theta)$, 
their Fourier transforms are $F_{_{_{\text{R}{\rightarrow}
\text{T}}}}(m){=}F_{_\text{T}}(m){=}\frac{1}{2\pi}\int_{-
\pi}^{\pi}\text{d}\theta\,e^{im\theta}$ and $F_{_{_{\text{T}
{\rightarrow}\text{R}}}}(\theta){=}F_{_\text{R}}(m){=}
\int_{-\pi}^{\pi}\text{d}\theta\;e^{im\theta}F_{_\text{R}}
(\theta)$. Next we apply the Fourier transformation on the master 
equations~(\ref{Eq:MasterEqs}). For example, the first master 
equation after Fourier transform reads
\begin{equation}
\begin{aligned}
&P_{\!_{t{+}\Delta t}}^{R}(k,\phi|m) = \vspace{1mm} \\
&\,(1{-}f_{_{_{\text{R}{\rightarrow}\text{T}}}}) 
\!\! \displaystyle \int \!\! \text{d}\alpha\,e^{im\alpha} 
\! \int \!\! d\gamma \, F_{_\text{R}}(\alpha{-}\gamma) \int \!\! 
\text{d}y \int \!\! \text{d}x \, e^{i\boldsymbol{k}\cdot
\boldsymbol r} \\
&P_{t}^{R}(x{-}v_{_\text{R}}\Delta t\cos\alpha,
y{-}v_{_\text{R}}\Delta t\sin\alpha|\gamma) \vspace{1mm} \\
&\,+ f_{_{_{\text{T}{\rightarrow}\text{R}}}} \int \!\! 
\text{d}\alpha \, e^{im\alpha} \! \int \!\! d\gamma \, F_{_{_{
\text{T}{\rightarrow}\text{R}}}}(\alpha{-}\gamma) 
\int \!\! \text{d}y \int \!\! \text{d}x \, e^{i\boldsymbol{k} 
\cdot\boldsymbol r} \\
&P_{t}^{T}(x{-}v_{_\text{R}}\Delta 
t\cos\alpha,y{-}v_{_\text{R}}\Delta t\sin\alpha|\gamma).
\end{aligned}
\label{Eq:MasterEqs-2}
\end{equation}
Then by using the $q$th order Bessel's function
\begin{equation}
J_q(z) = \frac{1}{2\pi i^{q}} \int_{-\pi}^{\pi} \!\! \text{d}\alpha \,\, 
e^{iz\cos\alpha} e^{-iq\alpha}, \nonumber
\end{equation}
replacing $e^{i k v_{_\text{R}} \Delta t \cos(\alpha{-}\phi)}$ 
with $\int_{\!{-}\pi}^{\pi} d\beta e^{i k v_{_\text{R}} \Delta 
t \cos\beta} \delta(\beta{-}(\alpha{-}\phi))$, and using
\begin{equation}
\delta(\beta{-}(\alpha{-}\phi))=\frac{1}{2\pi}\sum_{q{=}
-\infty}^{\infty} e^{-i q (\beta{-}(\alpha{-}\phi))}, \nonumber
\end{equation}
it follows that
\begin{equation}
\begin{aligned}
P_{\!_{t{+}\Delta t}}^{\text{R}}(k,\phi|m) = 
&\sum_{q=-\infty}^{\infty} \! i^q\,e^{-iq\phi} J_q(k \, 
v_{_\text{R}} \Delta t) \times \\
&\Big[(1{-}f_{_{_{\text{R}{\rightarrow}\text{T}}}})
\, F_{_\text{R}}(m{+}q)\,P_{t}^{\text{R}}
(k,\phi|m{+}q) \\
&+ f_{_{_{\text{T}{\rightarrow}\text{R}}}} \, 
F_{_\text{R}}(m{+}q)\, P_{t}^{\text{T}}(k,\phi|m{+}q)\Big].
\end{aligned}
\label{Eq:MasterEqs-3}
\end{equation}
$P_{\!_{t{+}\Delta t}}^{R}(k,\phi|m)$ can be expanded as a 
Taylor series
\begin{equation}
\begin{aligned}
P_{t{+}\Delta t}^R(k,\phi|m) &= Q_{0,t{+}\Delta t}^R(\phi|m) + 
i\,k\,v_{_\text{R}}\,\Delta t \, Q_{1,t{+}\Delta t}^R(\phi|m) \\
&-\frac{1}{2} k^2 \, v_{_\text{R}}^2 \, (\Delta t)^2 \, 
Q_{2,t{+}\Delta t}^R(\phi|m)+ \cdot \cdot \cdot.
\end{aligned}
\end{equation}
We expand both sides of Eq.~(\ref{Eq:MasterEqs-3}) and collect 
all terms with the same power in $k$. As a result, recursion 
relations for the Taylor expansion coefficients can be obtained. 
For instance, for the terms with power $0$ in $k$ one finds
\begin{equation}
\begin{aligned}
Q_{0,t{+}\Delta t}^{\text{R}}(\phi|m)= 
&(1{-}f_{_{_{\text{R}{\rightarrow}\text{T}}}}) F_{_\text{R}}(m) 
Q_{0,t}^{\text{R}}(\phi|m) \\
&+ f_{_{_{\text{T}{\rightarrow}\text{R}}}} 
\, F_{_\text{R}}(m) \, Q_{0,t}^{\text{T}}(\phi|m).
\end{aligned}
\end{equation}
Similarly, the expansion coefficients of terms with higher powers 
in $k$ can be calculated and the procedure is repeated for the 
second master equation in (\ref{Eq:MasterEqs}). As a result, a 
set of coupled equations is obtained for each expansion coefficient, 
connecting time steps $t{+}\Delta t$ and $t$. Applying a $z$-transform 
$Q(z){=}\sum_{t{=}0}^\infty Q_t z^{-t}$ enables one to solve these 
sets of equations. Particularly the coefficients of terms with 
power 2 in $k$, i.e.\ $Q_2^{\text{R}}(z,\phi|m)$ and $Q_2^{\text{T}}
(z,\phi|m)$, are useful to calculate the MSD
\begin{equation}
\langle x^2 \rangle(z) = (\Delta t)^2\,\Big(v_{_\text{R}}^2\,
Q_2^{\text{R}}(z,0|0){+}v_{_\text{T}}^2\,Q_2^{\text{T}}(z,0|0)\Big).
\end{equation}
Finally we obtain the following exact expression for the MSD in $z$ space
\begin{widetext}
\begin{equation}
\begin{aligned}
\langle x^2 \rangle(z) {=} &\Bigg[ \displaystyle\frac{z(1{-}f_{_{_{\text{R}
{\rightarrow}\text{T}}}}{-}f_{_{_{\text{T}{\rightarrow}\text{R}}}})
P_0^R}{z{-}1{+}f_{_{_{\text{R}{\rightarrow}\text{T}}}}{+}
f_{_{_{\text{T}{\rightarrow}\text{R}}}}}+\displaystyle\frac{z^2
\,f_{_{_{\text{T}{\rightarrow}\text{R}}}}}{G_0(z)} \Bigg] 
\Bigg[ \frac{z^2}{(z{-}1)G_1(z)} {-} \frac{1}{2(z{-}1)} \Bigg] 
v_{_\text{R}}^2 \, (\Delta t)^2 + \\
&\Bigg[ \displaystyle\frac{-z(1{-}f_{_{_{\text{R}
{\rightarrow}\text{T}}}}{-}f_{_{_{\text{T}{\rightarrow}\text{R}}}})
P_0^R}{z{-}1{+}f_{_{_{\text{R}{\rightarrow}\text{T}}}}{+}
f_{_{_{\text{T}{\rightarrow}\text{R}}}}}+\displaystyle\frac{z^2
(1{-}f_{_{_{\text{T}{\rightarrow}\text{R}}}})}{G_0(z)} 
-\displaystyle\frac{z(1{-}f_{_{_{\text{T}{\rightarrow}\text{R}}}}
{-}f_{_{_{\text{R}{\rightarrow}\text{T}}}})}{G_0(z)} \Bigg] \times \\
& \Bigg[ \frac{z\left[z{-}(1{-}f_{_{_{\text{R}{\rightarrow}\text{T}}}})
p\right]}{(z{-}1)G_1(z)} v^2_{_\text{T}} {+} \frac{z}{(z{-}1)G_1(z)} 
f_{_{_{\text{T}{\rightarrow}\text{R}}}} p \, v_{_\text{R}}  v_{_\text{T}} 
{-} \frac{1}{2(z{-}1)} v^2_{_\text{T}} \Bigg] (\Delta t)^2,
\end{aligned}
\label{Eq:MSD-2}
\end{equation}
\end{widetext}
where $G_0(z){=}(z{-}1)(z{-}1{+}f_{_{_{\text{T}{\rightarrow}\text{R}}}}{+}
f_{_{_{\text{R}{\rightarrow}\text{T}}}})$ and $G_1(z){=}z\big(z{-}(1{-}
f_{_{_{\text{R}{\rightarrow}\text{T}}}})\,p \big)$. By inverse 
$z$-transforming Eq.~(\ref{Eq:MSD-2}), the MSD can be obtained as 
a function of time. The resulting general expression for the MSD $\langle 
r^2 \rangle(t) {=}2\langle x^2 \rangle(t)$ is lengthy and depends 
on the run persistency $p$, the speeds $v_{_\text{R}}$ and $v_{_\text{T}}$, 
the transition probabilities $f_{_{_{\text{R}{\rightarrow}\text{T}}}}$ 
and $f_{_{_{\text{T}{\rightarrow}\text{R}}}}$, and the probability of 
initially starting in the run $P_0^R$ or tumble state $P_0^T{=}1{-}P_0^R$. 
$\langle r^2 \rangle(t)$ typically consists of linear and exponentially 
decaying terms with $t$ as well as time-independent terms, as shown in 
the following in the special case of constant velocity and the initial 
condition of starting in the run state. By choosing $\Delta t{=}1$, 
$v_{_\text{R}}{=}v_{_\text{T}}{=}1$, and the initial condition $P_0^R
{=}1$, the general expression of $\langle r^2 \rangle(t)$ reduces to 
\begin{widetext}
\begin{equation}
\begin{aligned}
\langle r^2 \rangle(t) {=} &\frac{\Big(p 
\big((f_{_{_{\text{T}{\rightarrow}\text{R}}}}{-}1) f_{_{_{\text{R}{\rightarrow}\text{T}}}}
{+}f_{_{_{\text{T}{\rightarrow}\text{R}}}}{+}f_{_{_{\text{R}{\rightarrow}\text{T}}}}^2\big)
{+}f_{_{_{\text{T}{\rightarrow}\text{R}}}}{+}f_{_{_{\text{R}{\rightarrow}\text{T}}}}\Big)}
{p\,(f_{_{_{\text{R}{\rightarrow}\text{T}}}}{-}1) (f_{_{_{\text{T}{\rightarrow}\text{R}}}}
{+}f_{_{_{\text{R}{\rightarrow}\text{T}}}}){+}f_{_{_{\text{T}{\rightarrow}\text{R}}}}{+}
f_{_{_{\text{R}{\rightarrow}\text{T}}}}} \,t - \\
&\frac{2\,p\,(f_{_{_{\text{R}{\rightarrow}\text{T}}}}{-}1) \big(f_{_{_{\text{R}{\rightarrow}\text{T}}}} 
p (f_{_{_{\text{T}{\rightarrow}\text{R}}}}{+}f_{_{_{\text{R}{\rightarrow}\text{T}}}}{-}2){+}
f_{_{_{\text{T}{\rightarrow}\text{R}}}}{+}f_{_{_{\text{R}{\rightarrow}\text{T}}}}{+}p{-}1\big)
\big(p(1{-}f_{_{_{\text{R}{\rightarrow}\text{T}}}})\big)^t}{\big(p(f_{_{_{\text{R}{\rightarrow}\text{T}}}}
{-}1){+}1\big)^2 (f_{_{_{\text{T}{\rightarrow}\text{R}}}}{-}f_{_{_{\text{R}{\rightarrow}\text{T}}}} 
p{+}f_{_{_{\text{R}{\rightarrow}\text{T}}}}{+}p{-}1)} + \\
&\frac{2\,p\,f_{_{_{\text{R}{\rightarrow}\text{T}}}} (1{-}f_{_{_{\text{T}{\rightarrow}\text{R}}}}
{-}f_{_{_{\text{R}{\rightarrow}\text{T}}}})^{t{+}2}}{(f_{_{_{\text{T}{\rightarrow}\text{R}}}}
{+}f_{_{_{\text{R}{\rightarrow}\text{T}}}})^2 (f_{_{_{\text{T}{\rightarrow}\text{R}}}}{-}
f_{_{_{\text{R}{\rightarrow}\text{T}}}} p{+}f_{_{_{\text{R}{\rightarrow}\text{T}}}}{+}p{-}1)}{+} \\
&\frac{\Big(p \big((f_{_{_{\text{T}{\rightarrow}\text{R}}}}{-}1) 
f_{_{_{\text{R}{\rightarrow}\text{T}}}}{+}f_{_{_{\text{T}{\rightarrow}\text{R}}}}
{+}f_{_{_{\text{R}{\rightarrow}\text{T}}}}^2\big){+}f_{_{_{\text{T}{\rightarrow}\text{R}}}}
{+}f_{_{_{\text{R}{\rightarrow}\text{T}}}}\Big)}{p\,(f_{_{_{\text{R}{\rightarrow}\text{T}}}}
{-}1) (f_{_{_{\text{T}{\rightarrow}\text{R}}}}{+}f_{_{_{\text{R}{\rightarrow}\text{T}}}})
{+}f_{_{_{\text{T}{\rightarrow}\text{R}}}}{+}f_{_{_{\text{R}{\rightarrow}\text{T}}}}}{-} \\
&\frac{2\,p\big((f_{_{_{\text{T}{\rightarrow}\text{R}}}}{+}
f_{_{_{\text{R}{\rightarrow}\text{T}}}})^2{-}f_{_{_{\text{R}{\rightarrow}\text{T}}}}\big)
{+}(f_{_{_{\text{T}{\rightarrow}\text{R}}}}{+}f_{_{_{\text{R}{\rightarrow}\text{T}}}}
)^2}{\big(p\,(f_{_{_{\text{R}{\rightarrow}\text{T}}}}{-}1) (f_{_{_{\text{T}{\rightarrow}\text{R}}}}
{+}f_{_{_{\text{R}{\rightarrow}\text{T}}}}){+}f_{_{_{\text{T}{\rightarrow}\text{R}}}}
{+}f_{_{_{\text{R}{\rightarrow}\text{T}}}}\big)^2}{+} \\
&\frac{p^2 (f_{_{_{\text{R}{\rightarrow}\text{T}}}}{-}1) 
\Big((f_{_{_{\text{T}{\rightarrow}\text{R}}}}{+}f_{_{_{\text{R}{\rightarrow}\text{T}}}}) 
\big(f_{_{_{\text{T}{\rightarrow}\text{R}}}} (f_{_{_{\text{R}{\rightarrow}\text{T}}}}{-}1)
{+}(f_{_{_{\text{R}{\rightarrow}\text{T}}}}{-}3) f_{_{_{\text{R}{\rightarrow}\text{T}}}}\big)
{+}2 f_{_{_{\text{R}{\rightarrow}\text{T}}}}\Big)}{\big(p\,(f_{_{_{\text{R}{\rightarrow}\text{T}}}}
{-}1)  (f_{_{_{\text{T}{\rightarrow}\text{R}}}}{+}f_{_{_{\text{R}{\rightarrow}\text{T}}}})
{+}f_{_{_{\text{T}{\rightarrow}\text{R}}}}{+}f_{_{_{\text{R}{\rightarrow}\text{T}}}}\big)^2}.
\end{aligned}
\label{Eq:MSD-3}
\end{equation}
\end{widetext}

\begin{figure*}[t]
\begin{center}
\includegraphics[width=12cm]{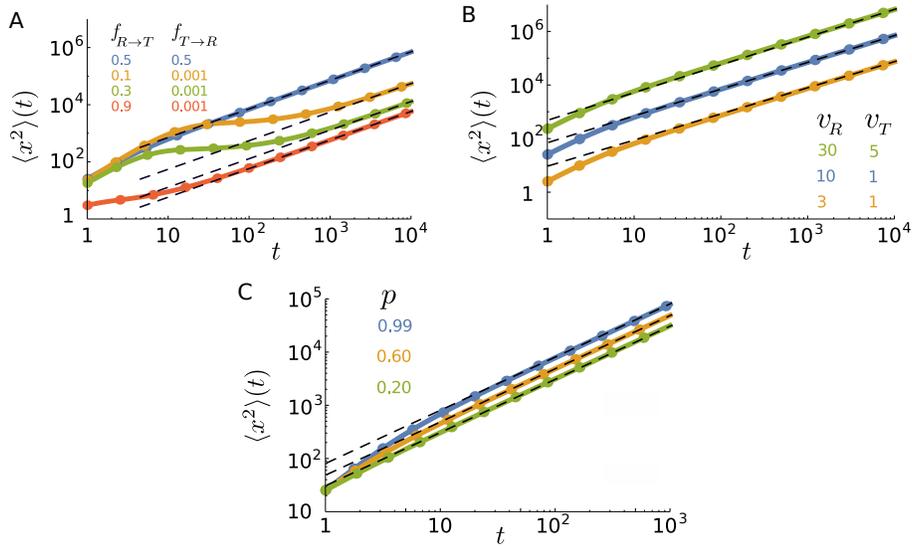}
\end{center}
\caption{Time evolution of the MSD for various (A) transition 
probabilities, (B) speeds, and (C) persistencies. The parameter 
values (unless varied) are taken to be $p{=}0.9$, $v_{_\text{R}}{=}10$, 
$v_{_\text{T}}{=}1$, $\Delta t{=}1$, $f_{_{_{\text{R}{\rightarrow}\text{T}}}}
{=}0.5$, $f_{_{_{\text{T}{\rightarrow}\text{R}}}}{=}0.5$ and $P_0^R{=}0.5$. 
The lines are obtained from the inverse z-transform of Eq.~(\ref{Eq:MSD-2}) 
and symbols denote simulation results. The dashed lines represent 
the asymptotic diffusion regime.}
\label{Fig2}
\end{figure*}

\begin{figure*}[t]
\begin{center}
\includegraphics[width=12cm]{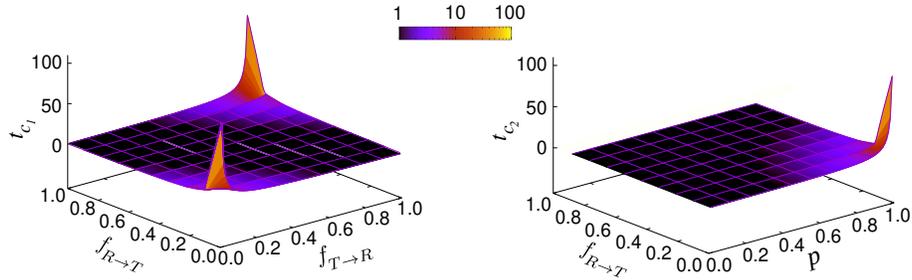}
\end{center}
\caption{Characteristic times $t_{c_{1}}$ and $t_{c_{2}}$ in terms of 
the transition probabilities $f_{_{_{\text{R}{\rightarrow}\text{T}}}}$ 
and $f_{_{_{\text{T}{\rightarrow}\text{R}}}}$ and run persistency $p$.}
\label{Fig3}
\end{figure*}

\section{Results and discussion}

We first investigate the time evolution of the MSD for different values of the 
key parameters $p$, $v_{_\text{R}}$, $v_{_\text{T}}$, $f_{_{_{\text{R}{\rightarrow}
\text{T}}}}$, $f_{_{_{\text{T}{\rightarrow}\text{R}}}}$, and $P_0^R$. As a simple 
check, the expression~(\ref{Eq:MSD-2}) for $f_{_{_{\text{R}{\rightarrow}\text{T}}}}{=}0$, 
$f_{_{_{\text{T}{\rightarrow}\text{R}}}}{=}1$, $v_{_\text{T}}{=}0$ and $P_0^R{=}1$ 
reduces to
\begin{equation}
\langle x^2 \rangle(z) {=} 
\frac{v_{_\text{R}}^2\,z\,(z{+}p)}{2(z{-}1)^2\,(z{-}p)}(\Delta t)^2,
\label{Eq:MSD-PRW-z-space}
\end{equation}
and by inverse $z$-transforming, the MSD for a single-state persistent random 
walk \citep{Nossal74,Tierno16}
\begin{equation}
\begin{aligned}
\langle x^2 \rangle(t) = \frac12 (\Delta t)^2 v^2_{_\text{R}} \Big[
\displaystyle\frac{1{+}p}{1{-}p}\,t+2p\displaystyle\frac{p^t
{-}1}{(1{-}p)^2} \Big]
\end{aligned}
\label{Eq:MSD-PRW}
\end{equation}
is recovered. In Fig.~\ref{Fig2}, we show how the MSD evolves in time 
for different values of the key parameters. We plot the general expression 
of $\langle x^2 \rangle(t)$, obtained from the inverse $z$-transforming of 
Eq.~(\ref{Eq:MSD-2}), and validate the analytical predictions by Monte 
Carlo simulations. A wide range of different types of anomalous dynamics 
can be observed on varying the parameters. While the short-time dynamics 
is typically superdiffusion (due to the combination of active and passive 
motion) and the long-term dynamics is diffusion in all cases, transitions 
between sub-, ordinary and super-diffusion occur on short and intermediate 
time scales. For some parameter values, the exponential terms of the MSD 
rapidly decay while the linear term is not yet big enough compared to the 
time-independent terms. In such a case, the constant terms dominate at 
intermediate time scales leading to the observed slow dynamics in this 
regime. The asymptotic dynamics is however diffusive since the linear 
term eventually dominates. It can also be seen that the crossover time 
to asymptotic diffusion varies by several orders of magnitude upon 
changing the parameter values. The crossover time can be characterized 
as the time at which the exponentially decreasing terms in $\langle x^2 
\rangle(t)$ become smaller than the terms which survive at long times. 
We find that the convergence of the MSD to its asymptotic diffusive 
form can be described by the sum of two exponential functions 
\begin{equation}
\langle x^2 \rangle(t){-}\langle x^2 \rangle(t{\rightarrow}\infty) \sim 
B_1\,\text{e}^{-t{/}t_{c_{1}}}+B_2\,\text{e}^{-t{/}t_{c_{2}}},
\label{Eq:Crossover}
\end{equation}
with the characteristic times $t_{c_{1}}{=}\displaystyle\frac{-1}{\ln|1{-}
f_{_{_{\text{R}{\rightarrow}\text{T}}}}{-}f_{_{_{\text{T}{\rightarrow}
\text{R}}}}|}$ and $t_{c_{2}}{=}\displaystyle\frac{-1}{\ln\big|p(1{-}
f_{_{_{\text{R}{\rightarrow}\text{T}}}})\big|}$. The prefactors $B_1$ 
and $B_2$ are functions of $p$, $v_{_\text{R}}$, $v_{_\text{T}}$, 
$f_{_{_{\text{R}{\rightarrow}\text{T}}}}$, $f_{_{_{\text{T}{\rightarrow}
\text{R}}}}$, and $P_0^R$. Figure~\ref{Fig3} shows how the characteristic 
times $t_{c_{1}}$ and $t_{c_{2}}$ vary upon changing the key parameters. 
Although the slopes of the exponential decays in Eq.~(\ref{Eq:Crossover}) 
are solely determined by the transition probabilities $f_{_{_{\text{R}{
\rightarrow}\text{T}}}}$ and $f_{_{_{\text{T}{\rightarrow}\text{R}}}}$ 
and the run persistency $p$, the crossover time to the asymptotic diffusive 
dynamics is also influenced by the other dynamic parameters of the model 
through the prefactors $B_1$ and $B_2$. For example, for the set of 
parameter values $p{=}0.9$, $v_{_\text{R}}{=}10$, $v_{_\text{T}}{=}0.1$, 
$f_{_{_{\text{R}{\rightarrow}\text{T}}}}{=}0.1$ and $f_{_{_{\text{T}{
\rightarrow}\text{R}}}}{=}0.01$, the convergence time (with $5\%$ accuracy) 
to the asymptotic dynamics for $P_0^R{=}1$ is nearly twice as long as for 
$P_0^R{=}0$.

\begin{figure*}[t]
\begin{center}
\includegraphics[width=12cm]{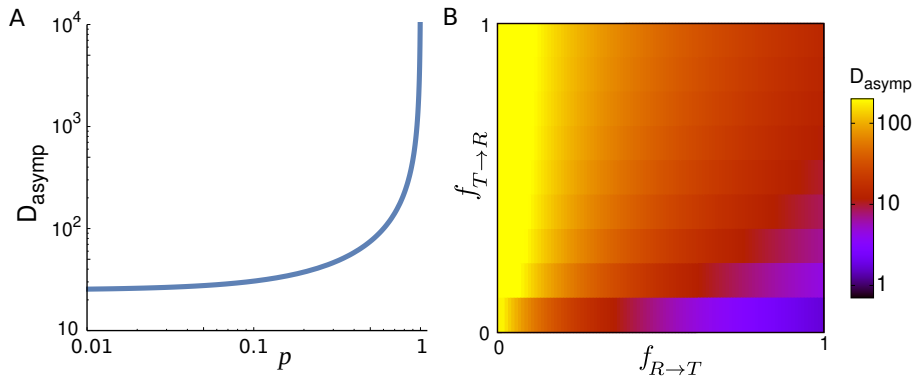}
\end{center}
\caption{(A) Asymptotic diffusion coefficient $D_\text{asymp}$ as a function 
of the run persistency $p$ for $v_{_\text{R}}{=}10$, $v_{_\text{T}}{=}1$, 
$\Delta t{=}1$, $f_{_{_{\text{R}{\rightarrow}\text{T}}}}{=}0.001$, and $f_{_{_{
\text{T}{\rightarrow}\text{R}}}}{=}0.9$. (B) $D_\text{asymp}$ in the space of 
transition probabilities $f_{_{_{\text{R}{\rightarrow}\text{T}}}}$ and $f_{_{_{
\text{T}{\rightarrow}\text{R}}}}$ for $v_{_\text{R}}{=}10$, $v_{_\text{T}}{=}1$, 
$\Delta t{=}1$, and $p{=}0.9$.}
\label{Fig4}
\end{figure*}

Figure~\ref{Fig2} also shows that the asymptotic diffusion constant 
$D_\text{asymp}$ varies by changing the key parameters. The differences 
in the $y$-intercept of the dashed (asymptotic) lines in log-log plots 
reflect the sensitivity of $D_\text{asymp}$ to the model parameters. 
By inverse z-transforming of Eq.~(\ref{Eq:MSD-2}) and taking the limit 
$t{\rightarrow}\infty$, we obtain $D_\text{asymp}$ (i.e.\ the coefficient 
of the term linear in time) in the general form as
\begin{widetext}
\begin{equation}
D_\text{asymp}= \frac14\,\Delta t\,
\frac{2 f_{_{_{\text{T}{\rightarrow}\text{R}}}} 
f_{_{_{\text{R}{\rightarrow}\text{T}}}} p\,v_{_\text{T}}\, 
v_{_\text{R}}{+}f_{_{_{\text{T}{\rightarrow}\text{R}}}} 
v_{_\text{R}}^2 \big(1{+}p(1{-}f_{_{_{\text{R}{\rightarrow}\text{T}}}} 
)\big){+}f_{_{_{\text{R}{\rightarrow}\text{T}}}} 
v_{_\text{T}}^2 \big(1{-}p(1{-}f_{_{_{\text{R}{\rightarrow}\text{T}}}}) 
\big)}{(f_{_{_{\text{T}{\rightarrow}\text{R}}}}{+}
f_{_{_{\text{R}{\rightarrow}\text{T}}}}) 
\big(1{-}p(1{-}f_{_{_{\text{R}{\rightarrow}\text{T}}}})\big)}.
\label{Eq:Dasymp}
\end{equation}
\end{widetext}
While the diffusion coefficient trivially increases with the speed, 
its dependency on $f_{_{_{\text{R}{\rightarrow}\text{T}}}}$, 
$f_{_{_{\text{T}{\rightarrow}\text{R}}}}$ and $p$ is more complicated 
and shown in Fig.~\ref{Fig4}. $D_\text{asymp}$ varies by several 
orders of magnitude as a function of these parameters. Under the specific 
condition $F_{_{_{\text{R}{\rightarrow}\text{T}}}}(\theta){=}
F_{_{_{\text{T}}}}(\theta){=}\delta(\theta)$ and $F_{_{_{\text{T}{
\rightarrow}\text{R}}}}(\theta){=}F_{_{_{\text{R}}}}(\theta)$ and 
$v_{_\text{T}}{=}0$, the walker stops when entering the tumble 
phase without changing its arrival direction and it returns smoothly 
to the run phase without experiencing a kick (i.e.\ a sharp change 
in the direction of motion). Motor-driven transport along cytoskeletal 
filaments in crowded cytoplasm exhibits such a run-and-pause dynamics 
\citep{Hafner16,Song18}. In this case, one obtains
\begin{equation}
\begin{aligned}
D_\text{asymp}^\text{run-pause}=\frac14\,\Delta t\,v^2_{_\text{R}} \frac{1{+}p}{1{-}p} 
\frac{f_{_{_{\text{T}{\rightarrow}\text{R}}}}}{f_{_{_{\text{T}
{\rightarrow}\text{R}}}}{+}f_{_{_{\text{R}{\rightarrow}\text{T}}}}}. 
\end{aligned}
\label{Eq:Dasymp-RunPause}
\end{equation}
In the limit $p{\rightarrow}1$ the trajectory becomes nearly straight 
implying that the randomization time and the covered area until reaching 
the asymptotic diffusive regime (and thus $D_\text{asymp}$) diverge.

\begin{figure}[t]
\begin{center}
\includegraphics[width=8cm]{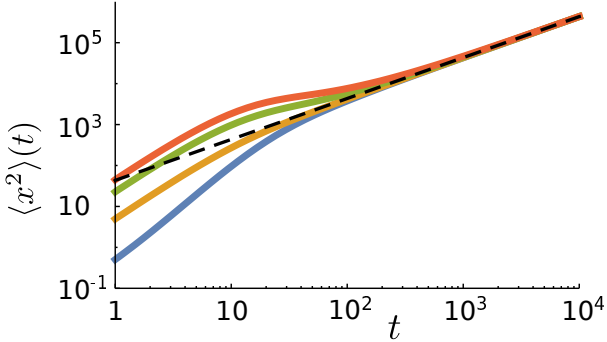}
\end{center}
\caption{The mean square displacement as a function of time for different 
values of the probability $P_0^R$ of initially starting in the run state 
in simulations (from top to bottom: $P_0^R{=}1.0, 0.5, 0.1, 0.0$). 
Other parameter values: $p{=}0.9$, $v_{_\text{R}}{=}10$, $v_{_\text{T}}{=}0.1$, 
$f_{_{_{\text{R}{\rightarrow}\text{T}}}}{=}0.1$ and $f_{_{_{\text{T}{\rightarrow}
\text{R}}}}{=}0.01$. The dashed line represents the analytical 
prediction via Eq.~(\ref{Eq:AsympDynamics}) for the same parameter values.}
\label{Fig5}
\end{figure}

Interestingly, $D_\text{asymp}$ in Eq.~(\ref{Eq:Dasymp}) is independent 
of $P_0^R$ and $P_0^T$, i.e.\ the initial condition of starting the motion 
in the run or tumble state. Thus the analytical results predict that 
the asymptotic diffusive dynamics, characterized by the linear time-dependence  
\begin{equation}
\langle x^2 \rangle(t{\rightarrow}\infty) = 2 D_\text{asymp} \, t,
\label{Eq:AsympDynamics}
\end{equation}
does not depend on the initial conditions. In Fig.~\ref{Fig5} we present the simulation 
results for several values of $P_0^R$. At long times, all curves merge and follow the 
analytical prediction Eq.~(\ref{Eq:AsympDynamics}). Note that only the linear term in 
time is independent of $P_0^R$ and the exponentially decaying and time-independent 
terms in the MSD depend on the initial conditions [see e.g. Eq.~(\ref{Eq:MSD-3})]. 
Although the process is Markovian it keeps initially for some time its memory of the 
initial direction and state of motion. However, the influence of the $P_0^R$-dependent 
terms vanishes in the limit $t{\rightarrow}\infty$ and the time dependence of the 
MSD appoaches the asymptotic linear form Eq.~(\ref{Eq:AsympDynamics}). 

\begin{figure*}[t]
\begin{center}
\includegraphics[width=12cm]{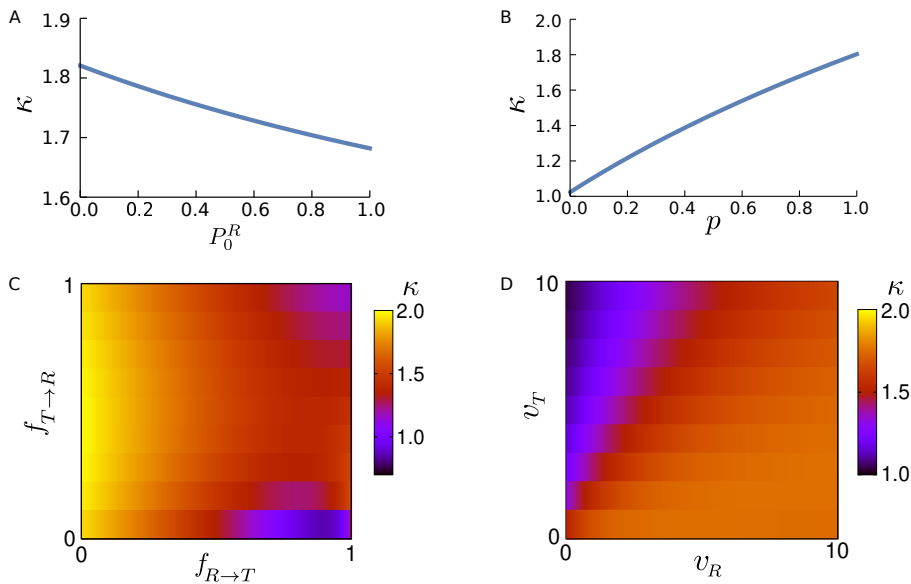}
\end{center}
\caption{The anomalous exponent $\kappa$ versus (A) the initial condition 
of motion $P_0^R$, (B) run persistency $p$, (C) transition probabilities 
$f_{_{_{\text{R}{\rightarrow}\text{T}}}}$ and $f_{_{_{\text{T}{\rightarrow}
\text{R}}}}$, and (D) speeds $v_{_\text{R}}$ and $v_{_\text{T}}$ via 
Eq.~(\ref{Eq:Exponent2}). The parameter values (unless varied) are taken to 
be $p{=}0.9$, $v_{_\text{R}}{=}10$, $v_{_\text{T}}{=}1$, $\Delta t{=}1$, 
$f_{_{_{\text{R}{\rightarrow}\text{T}}}}{=}0.3$, $f_{_{_{\text{T}{\rightarrow}
\text{R}}}}{=}0.5$ and $P_0^R{=}0.5$.}
\label{Fig6}
\end{figure*}

The short time dynamics is, however, strongly influenced by the choice 
of the initial conditions of motion. Figure~\ref{Fig5} shows that the 
initial slope of the MSD curve varies with $P_0^R$. One can assign an 
initial anomalous exponent $\kappa$ to the MSD curve by fitting the 
power-law $\langle r^2 \rangle {\sim} t^{\kappa}$. By choosing the first 
two data points of the MSD curve, the fitting leads to $\langle 
r^2 \rangle(t{=}2){/}\langle r^2 \rangle(t{=}1){=}2^\kappa$. 
Thus, the initial anomalous exponent $\kappa$ can be deduced from 
the MSD at $t{=}1,\,2$ as
\begin{equation}
\kappa=\ln\Bigg[\displaystyle\frac{\langle r^2 \rangle(t{=}2)}{\langle 
r^2 \rangle(t{=}1)}\Bigg]{/}\ln2,
\label{Eq:Exponent1}
\end{equation}
After replacing the MSD at $t{=}1,\,2$ obtained from Eq.~(\ref{Eq:MSD-2}) we get
\begin{widetext}
\begin{equation}
\begin{aligned}
\kappa= \ln\Bigg[\Big[&\Big(2 {-} 2 P_0^R {+} (3 {-} f_{_{_{\text{T}{\rightarrow}\text{R}}}} {-} 
f_{_{_{\text{R}{\rightarrow}\text{T}}}}) \big(f_{_{_{\text{T}{\rightarrow}\text{R}}}} 
(P_0^R{-}1) {+} f_{_{_{\text{R}{\rightarrow}\text{T}}}} P_0^R\big)\Big) v_{_\text{T}}^2 {+} \\
&2 f_{_{_{\text{T}{\rightarrow}\text{R}}}} \big(1 {-} 
f_{_{_{\text{T}{\rightarrow}\text{R}}}} {+} (f_{_{_{\text{T}{\rightarrow}\text{R}}}} 
{+} f_{_{_{\text{R}{\rightarrow}\text{T}}}}{-1}) P_0^R\big)\,p\,v_{_\text{T}}\,v_{_\text{R}} {+} \\
&\Big(2 P_0^R {+} (-3 {+} f_{_{_{\text{T}{\rightarrow}\text{R}}}} {+} 
f_{_{_{\text{R}{\rightarrow}\text{T}}}}) \big(f_{_{_{\text{T}{\rightarrow}\text{R}}}} 
(P_0^R{-}1) {+} f_{_{_{\text{R}{\rightarrow}\text{T}}}} P_0^R\big) {+} \\
&2 (f_{_{_{\text{R}{\rightarrow}\text{T}}}}{-}1) \big(-f_{_{_{\text{T}{\rightarrow}\text{R}}}} 
{+} (f_{_{_{\text{T}{\rightarrow}\text{R}}}} {+} f_{_{_{\text{R}{\rightarrow}\text{T}}}}{-}1) 
P_0^R\big) p\Big) v_{_\text{R}}^2\Big] / \\
&\hspace{-2.5mm}\Big[\big(1 {-} f_{_{_{\text{T}{\rightarrow}\text{R}}}} {-} 
(1{-}f_{_{_{\text{T}{\rightarrow}\text{R}}}} {-} f_{_{_{\text{R}{\rightarrow}\text{T}}}}) 
P_0^R\big) v_{_\text{T}}^2 {+} \big(f_{_{_{\text{T}{\rightarrow}\text{R}}}} {+} (1{-}
f_{_{_{\text{T}{\rightarrow}\text{R}}}} {-} f_{_{_{\text{R}{\rightarrow}\text{T}}}}) 
P_0^R\big) v_{_\text{R}}^2\Big]\Bigg]/\ln2.
\end{aligned}   
\label{Eq:Exponent2}
\end{equation}
\end{widetext}
Figure~\ref{Fig6}(A) shows how the initial conditions of motion influences the 
initial anomalous exponent for a given set of parameters. Note that the displayed 
monotonic growth of $\kappa$ with $P_0^R$ does not hold in general, as we observed 
decreasing as well as nonmonotonic functionalities by varying other parameter 
values. However, $\kappa$ increases monotonically with $p$ in all parameter 
regimes as shown in Fig.~\ref{Fig6}(B). Moreover, figures~\ref{Fig6}(C) and 
(D) show that $\kappa$ also varies widely with the speed and transition 
probabilities. Because of combining an active run state ($0{<}p{<}1$) and normal 
diffusion (tumble state), $\kappa$ remains above $1$ (superdiffusion). However, 
by generalizing the run state to include subdiffusive motion (i.e.\ when 
$-1{<}p{<}1$), $\kappa$ can decrease below 1.

To better understand the role of the initial conditions, we note that the 
steady probabilities $P^R_\text{steady}$ and $P^T_\text{steady}$ of finding 
the particle in each of the two states is determined by the transition 
probabilities $f_{_{_{\text{R}{\rightarrow}\text{T}}}}$ and 
$f_{_{_{\text{T}{\rightarrow}\text{R}}}}$. Therefore, the influence of the 
initial condition of starting the motion in any of the two states gradually 
vanishes as the probabilities $P^R(t)$ and $P^T(t)$ of finding the particle 
in the run or tumble state gradually approach their steady values. By 
considering a discrete time Markov process with transition probabilities 
$f_{_{_{\text{R}{\rightarrow}\text{T}}}}$ and $f_{_{_{\text{T}{\rightarrow}
\text{R}}}}$, the probabilities at time $t$ can be obtained from those at 
time $t{-}1$ as
\begin{equation}
\Big(P^R(t), P^T(t)\Big)=\big(P^R(t{-}1), P^T(t{-}1)\Big)
\Big[\!\!\!
\begin{array}{ccc}
1{-}f_{_{_{\text{R}{\rightarrow}\text{T}}}} & f_{_{_{\text{R}{\rightarrow}\text{T}}}} \\
f_{_{_{\text{T}{\rightarrow}\text{R}}}} & 1{-}f_{_{_{\text{T}{\rightarrow}\text{R}}}}
\end{array} 
\!\!\!\Big].
\end{equation}
By applying this relation recursively, one can derive the probabilities at time 
$t$ based on the initial probabilities
\begin{widetext}
\begin{equation}
\begin{aligned} 
\Big(P^R(t), &P^T(t)\Big)=\Big(P_0^R, P_0^T\Big) 
\Big[\!\!\!
\begin{array}{ccc}
1{-}f_{_{_{\text{R}{\rightarrow}\text{T}}}} & f_{_{_{\text{R}{\rightarrow}\text{T}}}} \\
f_{_{_{\text{T}{\rightarrow}\text{R}}}} & 1{-}f_{_{_{\text{T}{\rightarrow}\text{R}}}}
\end{array} 
\!\!\!\Big]^t \\
&=\Big(P_0^R, P_0^T\Big)\,\frac{1}{f_{_{_{\text{R}{\rightarrow}\text{T}}}}{+}
f_{_{_{\text{T}{\rightarrow}\text{R}}}}}\,\Bigg[\!\!\!
\begin{array}{ccc}
f_{_{_{\text{T}{\rightarrow}\text{R}}}}{+}f_{_{_{\text{R}{\rightarrow}\text{T}}}}
(1{-}f_{_{_{\text{T}{\rightarrow}\text{R}}}}{-}f_{_{_{\text{R}{\rightarrow}\text{T}}}})^t 
& f_{_{_{\text{R}{\rightarrow}\text{T}}}} \Big(1{-}(1{-}f_{_{_{\text{T}{\rightarrow}\text{R}}}}
{-}f_{_{_{\text{R}{\rightarrow}\text{T}}}})^t\Big) \\
f_{_{_{\text{T}{\rightarrow}\text{R}}}} \Big(1{-}(1{-}f_{_{_{\text{T}{\rightarrow}\text{R}}}}
{-}f_{_{_{\text{R}{\rightarrow}\text{T}}}})^t\Big) 
& f_{_{_{\text{R}{\rightarrow}\text{T}}}}{+}f_{_{_{\text{T}{\rightarrow}\text{R}}}}
(1{-}f_{_{_{\text{T}{\rightarrow}\text{R}}}}{-}f_{_{_{\text{R}{\rightarrow}\text{T}}}})^t
\end{array} 
\!\!\!\Bigg].
\end{aligned} 
\end{equation}
Thus the evolution of $P^R(t)$ and $P^T(t)$ obeys
\begin{align}
P^R(t) &= \frac{f_{_{_{\text{T}{\rightarrow}\text{R}}}}}{f_{_{_{\text{T}
{\rightarrow}\text{R}}}}+f_{_{_{\text{R}{\rightarrow}\text{T}}}}} 
+ \frac{\left(1-f_{_{_{\text{T}{\rightarrow}\text{R}}}}-f_{_{_{\text{R}
{\rightarrow}\text{T}}}}\right)^t}{f_{_{_{\text{T}{\rightarrow}\text{R}}}}
+f_{_{_{\text{R}{\rightarrow}\text{T}}}}}\left(f_{_{_{\text{R}{\rightarrow}
\text{T}}}} P_0^R-f_{_{_{\text{T}{\rightarrow}\text{R}}}}(1
-P_0^R)\right), \notag \\
P^T(t) &= \frac{f_{_{_{\text{R}{\rightarrow}\text{T}}}}}{f_{_{_{\text{T}
{\rightarrow}\text{R}}}}+f_{_{_{\text{R}{\rightarrow}\text{T}}}}} 
- \frac{\left(1-f_{_{_{\text{T}{\rightarrow}\text{R}}}}-f_{_{_{\text{R}
{\rightarrow}\text{T}}}}\right)^t}{
f_{_{_{\text{T}{\rightarrow}\text{R}}}}+f_{_{_{\text{R}{\rightarrow}
\text{T}}}}}\left(f_{_{_{\text{R}{\rightarrow}\text{T}}}} P_0^R - 
f_{_{_{\text{T}{\rightarrow}\text{R}}}}(1
-P_0^R)\right), 
\label{Eq:MarkovP}
\end{align}
\end{widetext}
leading to the steady probabilities $P^R_\text{steady}{=}
\frac{f_{_{_{\text{T}{\rightarrow}\text{R}}}}}{f_{_{_{\text{T}
{\rightarrow}\text{R}}}}{+}f_{_{_{\text{R}{\rightarrow}\text{T}}}}}$ 
and $P^T_\text{steady}{=}\frac{f_{_{_{\text{R}{\rightarrow}
\text{T}}}}}{f_{_{_{\text{T}{\rightarrow}\text{R}}}}{+}
f_{_{_{\text{R}{\rightarrow}\text{T}}}}}$. If one starts with the 
initial condition $P_0^R{=}P^R_\text{steady}$, the system is 
immediately equilibrated. Otherwise, the choice of the initial 
conditions of motion affects the short-time dynamics and diversifies 
the transient anomalous diffusive regimes. According 
to Eq.~(\ref{Eq:MarkovP}), the relaxation of the probabilities 
toward their steady values follow an exponential decay $P^R(t),\,
P^T(t)\sim\text{e}^{-t{/}t_{m}}$ with $t_{m}{=}
\displaystyle\frac{-1}{\ln|1{-}f_{_{_{\text{R}{\rightarrow}
\text{T}}}}{-}f_{_{_{\text{T}{\rightarrow}\text{R}}}}|}$. While 
the characteristic time for the relaxation of the probabilities 
solely depends on the transition probabilities, the characteristic 
time for the crossover to asymptotic dynamics is influenced additionally 
by the run persistency, as we showed previously in Eq.~(\ref{Eq:Crossover}). 
Therefore, there are two independent relaxation times $t_{m}({=}t_{c_{1}})$ 
and $t_{c_{2}}$. In case the relaxations occur on different time scales, 
two distinct crossovers in the time evolution of the MSD may be observed 
in general as can be seen in Fig.~\ref{Fig2}(A).

\section{Conclusion}

We presented a persistent random walk model to study the stochastic 
dynamics of particles with active fast and passive slow motility 
modes. We derived an exact analytical expression for the mean 
square displacement, which allows one to analyze the transient 
anomalous transport regimes on short time scales and also extract 
the characteristics of the asymptotic diffusive motion such as the 
crossover time and the long-term diffusion constant. In particular 
we showed that while the choice of the initial conditions of motion 
influences the anomalous diffusion at short times, the asymptotic 
behavior remains independent of it and is entirely controlled by 
the run persistency, the velocities of the run and tumble states and 
the transition probabilities between the two states.

This work was financially supported by the German Research Foundation 
(DFG) within the Collaborative Research Center SFB 1027 (A3, A7).

\bibliography{Refs}

\end{document}